\documentclass[prl,preprint,amsmath,amssymb,superscriptaddress]{revtex4-2}
\usepackage{epsfig}
\usepackage{graphicx}
\usepackage{hyperref}
\usepackage{color}
\usepackage{bm}
\usepackage{upgreek}
\usepackage{soul}

\begin{document}

\renewcommand{\figurename}{Fig.}
\renewcommand{\tablename}{Tab.}

\title{Unconventional relativistic spin polarization of electronic bands in an altermagnet}

\author{A.~Dal~Din}
\thanks{These authors contributed equally}
\affiliation{School of Physics and Astronomy, University of Nottingham, Nottingham NG7 2RD, United Kingdom}

\author{D.A.~Usanov}
\thanks{These authors contributed equally}
\affiliation{Institut de Physique, \'{E}cole Polytechnique F\'{e}d\'{e}rale de Lausanne, CH-1015 Lausanne, Switzerland}
\affiliation{Photon Science Division, Paul Scherrer Institut, CH-5232 Villigen, Switzerland}

\author{L.~\v{S}mejkal}
\thanks{These authors contributed equally}
\email{Corresponding authors: lsmejkal@pks.mpg.de, jminar@ntc.zcu.cz, hugo.dil@epfl.ch, jungw@fzu.cz, Peter.Wadley@nottingham.ac.uk}
\affiliation{Max Planck Institute for the Physics of Complex Systems, 01187 Dresden, Germany}
\affiliation{Max Planck Institute for Chemical Physics of Solids, 01187 Dresden, Germany} 
\affiliation{Institute of Physics, Czech Academy of Sciences, Cukrovarnick\'{a} 10, 162 00 Praha 6 Czech Republic}

\author{S.~W.~D'Souza}
\thanks{These authors contributed equally}
\affiliation{University of West Bohemia, New Technologies Research Center, Plzen 30100, Czech Republic}

\author{F.~Guo}
\affiliation{Institut de Physique, \'{E}cole Polytechnique F\'{e}d\'{e}rale de Lausanne, CH-1015 Lausanne, Switzerland}

\author{O.~J.~Amin}
\affiliation{School of Physics and Astronomy, University of Nottingham, Nottingham NG7 2RD, United Kingdom}
\author{E.~M.~Dawa}
\affiliation{School of Physics and Astronomy, University of Nottingham, Nottingham NG7 2RD, United Kingdom}
\author{R.~P.~Campion}
\affiliation{School of Physics and Astronomy, University of Nottingham, Nottingham NG7 2RD, United Kingdom}
\author{K.~W.~Edmonds}
\affiliation{School of Physics and Astronomy, University of Nottingham, Nottingham NG7 2RD, United Kingdom}
\author{B.~Kiraly}
\affiliation{School of Physics and Astronomy, University of Nottingham, Nottingham NG7 2RD, United Kingdom}
\author{A.~W.~Rushforth}
\affiliation{School of Physics and Astronomy, University of Nottingham, Nottingham NG7 2RD, United Kingdom}

\author{C.~Polley}
\affiliation{MAX IV Laboratory, Lund University, 221 00, Lund, Sweden}

\author{M.~Leandersson}
\affiliation{MAX IV Laboratory, Lund University, 221 00, Lund, Sweden}

\author{E.~Golias}
\affiliation{MAX IV Laboratory, Lund University, 221 00, Lund, Sweden}

\author{Y.~Niu}
\affiliation{MAX IV Laboratory, Lund University, 221 00, Lund, Sweden}

\author{S.~Telkamp}
\affiliation{Solid State Physics Laboratory, ETH Zürich, CH-8093 Zürich, Switzerland} 

\author{F.~Krizek}
\affiliation{Institute of Physics, Czech Academy of Sciences, Cukrovarnick\'{a} 10, 162 00 Praha 6 Czech Republic}

\author{A.~Birk~Hellenes}
\affiliation{Institute of Physics, Czech Academy of Sciences, Cukrovarnick\'{a} 10, 162 00 Praha 6 Czech Republic}
\affiliation{Institute of Physics, Johannes Gutenberg University Mainz, Staudingerweg 9, D-55099 Mainz, Germany}

\author{J.~Priessnitz}
\affiliation{Max Planck Institute for the Physics of Complex Systems, 01187 Dresden, Germany}

\author{W.~H.~Campos}
\affiliation{Max Planck Institute for the Physics of Complex Systems, 01187 Dresden, Germany}

\author{J.~Krempask\'y}
\affiliation{Photon Science Division, Paul Scherrer Institut, CH-5232 Villigen, Switzerland}

\author{J.~Min\'ar}
\email{Corresponding authors: lsmejkal@pks.mpg.de, jminar@ntc.zcu.cz, hugo.dil@epfl.ch, jungw@fzu.cz, Peter.Wadley@nottingham.ac.uk}
\affiliation{University of West Bohemia, New Technologies Research Center, Plzen 30100, Czech Republic}

\author{T.~Jungwirth}
\email{Corresponding authors: lsmejkal@pks.mpg.de, jminar@ntc.zcu.cz, hugo.dil@epfl.ch, jungw@fzu.cz, Peter.Wadley@nottingham.ac.uk}
\affiliation{Institute of Physics, Czech Academy of Sciences, Cukrovarnick\'{a} 10, 162 00 Praha 6 Czech Republic}
\affiliation{School of Physics and Astronomy, University of Nottingham, Nottingham NG7 2RD, United Kingdom}

\author{ J.~H.~Dil}
\email{Corresponding authors: lsmejkal@pks.mpg.de, jminar@ntc.zcu.cz, hugo.dil@epfl.ch, jungw@fzu.cz, Peter.Wadley@nottingham.ac.uk}
\affiliation{Institut de Physique, \'{E}cole Polytechnique F\'{e}d\'{e}rale de Lausanne, CH-1015 Lausanne, Switzerland}
\affiliation{Photon Science Division, Paul Scherrer Institut, CH-5232 Villigen, Switzerland}

\author{P.~Wadley}
\email{Corresponding authors: lsmejkal@pks.mpg.de, jminar@ntc.zcu.cz, hugo.dil@epfl.ch, jungw@fzu.cz, Peter.Wadley@nottingham.ac.uk}
\affiliation{School of Physics and Astronomy, University of Nottingham, Nottingham NG7 2RD, United Kingdom}

\date{\today}

\begin{abstract}
Altermagnetism is a recently identified phase with a $d$, $g$ or $i$-wave spin symmetry of magnetic ordering.  Its discovery opens new research fronts at intersections of magnetism and spintronics with fields ranging from superconductivity to topological and relativistic quantum physics. Here we demonstrate an unconventional relativistic spin polarization in an altermagnet by spin and angle resolved photoemission spectroscopy of electronic bands in single-domain MnTe. The relativistic spin-orbit coupling origin is revealed by observing that the alternating momentum-dependent spin polarization is orthogonal to the magnetic-ordering vector. The collinearity, even-parity and time-reversal-odd  nature of the demonstrated relativistic spin polarization in the altermagnet is unparalleled in conventional forms of the relativistic spin polarization. Our experimental results and methodology are supported by non-relativistic spin-symmetry and relativistic magnetic-symmetry analyses, and microscopic ab initio ground-state and photoemission theory. 
\end{abstract} 

\maketitle

The relativistic spin polarization of electronic states enables a range of  phenomena explored and exploited, among others, in the fields of topological quantum matter  or spintronics \cite{Nagaosa2010,Franz2013,Sinova2015,Bradlyn2017,Smejkal2018,Zang2018,Manchon2019,Vergniory2019,Tokura2019,Xu2020,Elcoro2021,Smejkal2022AHEReview}. In the normal phases of electronic systems, which preserve symmetries of the Hamiltonian, the phenomenology of the relativistic spin polarization is well established. In the non-relativistic limit, the Hamiltonian of the electronic system has the spin-space rotation symmetry which implies that the entire non-relativistic electronic structure is spin unpolarized  in the normal phase. The relativistic spin polarization occurs due to the spin-orbit coupling term  $\sim 1/c^2\,{\bf s}\cdot({\bf k} \times{\bf E})$ in the relativistic Hamiltonian \cite{Strange1998}, which breaks the spin-space rotation symmetry. Here $c$ is the speed of light, {\bf s} denotes spin, {\bf k} momentum and {\bf E} electric field. The time-reversal symmetry of the relativistic Hamiltonian implies that  the electronic states in the normal phase can be spin polarized only if the Hamiltonian of the system breaks inversion symmetry.  (Recall that  electronic states are Kramers spin degenerate \cite{Kramers1930,Wigner1932} in the entire Brillouin zone in the presence of the symmetry combining time-reversal and inversion.) The conventional  relativistic spin polarization in a normal phase is thus odd-parity and time-reversal-even, and  the axis of the spin polarization tends to vary with momentum forming a non-collinear momentum-dependent spin texture \cite{Strange1998,Winkler2003}. 

Magnetically ordered ground states spontaneously break the spin-space rotation symmetry, which can lead to the spin polarization of the electronic states already in the non-relativistic limit. The remarkable feature of the $d$, $g$ or $i$-wave altermagnetic order is a retained symmetry combining the spin-space rotation with a real-space rotation (proper or improper, and symmorphic or non-symmorphic) \cite{Smejkal2021a,Smejkal2022a}. Reminiscent of the normal phases, this non-relativistic spin symmetry enforces unpolarized electronic states. However, unlike the normal phases, the symmetry-enforced unpolarized states in the non-relativistic electronic structure are limited to 2, 4, or 6 nodal surfaces in the Brillouin zone of the  ordered $d$, $g$ or $i$-wave altermagnetic phases, respectively  \cite{Smejkal2021a,Smejkal2022a}. 

In Fig.~1 we highlight the $k_z=0$ nodal plane in the momentum space of a $g$-wave altermagnet MnTe \cite{Smejkal2021a,Betancourt2021,Mazin2023,Hariki2023,Krempasky2024,Lee2024,Osumi2024,Hajlaoui2024,Amin2024}. This nodal plane reflects the non-relativistic spin symmetry $[C_2||M_z]$ (see Fig.~1(a)) of the altermagnetic order, combining the spin-space rotation  ($C_2$) with a real-space $z$-plane mirror transformation ($M_z$) \cite{Smejkal2021a}. For the  Mn magnetic moments aligned in the $z$-plane, our ab initio density-functional-theory (DFT, for details see Methods) calculations in Fig.~1b,c show a sizable relativistic spin splitting  on the $k_z=0$ nodal plane, with the spin polarization pointing along the out-of-plane $z$-axis. Remarkably, the relativistic spin polarization occurs despite the inversion symmetry of the magnetic MnTe crystal. Here the lifting of the Kramers spin degeneracy is allowed due to the broken time-reversal symmetry by the altermagnetic order \cite{Smejkal2020,Smejkal2021a}. As a result, the relativistic spin polarization on the nodal plane is even-parity and time-reversal-odd. This, in combination with the collinear spin polarization (momentum-independent spin-polarization axis), shown in Fig.~1(c), is in striking contrast to the conventional relativistic spin polarization in normal phases with broken inversion symmetry. In the latter case, the spin polarization tends to form non-collinear spin textures in the momentum space (e.g. Rashba spin texture), has odd parity, and is time-reversal-even.

The absence of in-plane components of the relativistic spin polarization on the $k_z=0$ nodal plane is due to the relativistic magnetic mirror symmetry ${\cal M}_z$, highlighted in Fig.~1(a). (Note that the relativistic ${\cal M}_z$ transformation acts simultaneously in the real space and the spin space, and flips the in-plane Mn magnetic moments which are parallel to the mirror plane.)  As a result,  only the magnitude or the sign of the relativistic spin polarization can vary with the momentum, while the polarization axis is fixed along the $z$-direction for all momenta on the $k_z=0$ nodal plane.  This is highlighted in Fig.~1(c) by the calculated expectation value of spin on top of the $k_z=0$ valence band structure. For the  N\'eel vector along the $[1\bar{1}00]$ magnetic easy axis of MnTe, which corresponds to the $\boldsymbol\Gamma - {\bf M}$ direction in the Brillouin zone, the spin polarization has one sign for the $\boldsymbol\Gamma - {\bf K}_1$ direction, which is orthogonal to the $\boldsymbol\Gamma - {\bf M}$ direction, and the opposite sign for the $\boldsymbol\Gamma - {\bf K}_2$ direction, rotated by $60^\circ$ from the $\boldsymbol\Gamma - {\bf M}$ direction.  

The above unconventional phenomenology of the relativistic spin polarization  in the altermagnet is unparalleled, besides the normal phases, also among other magnetically ordered phases. For comparison, conventional ferromagnets have no spin-symmetry protected nodal surfaces in their non-relativistic electronic structure. Conventional antiferromagnets with opposite magnetic moments related by translation have a time-reversal-even electronic structure, and with opposite magnetic moments related by inversion have a spin-unpolarized electronic structure over the entire Brillouin zone, without or with the relativistic spin-orbit coupling \cite{Smejkal2021a,Smejkal2022a}. 

We now experimentally demonstrate the unconventional phenomenology of the relativistic altermagnetic spin polarization in our single-domain samples of altermagnetic MnTe  by spin and angle resolved photoemission spectroscopy. Thin films of $\alpha$-MnTe(0001) were grown by molecular-beam epitaxy  on an InP(111) substrate. The high single-crystal quality of the films is evidenced by transmission electron microscopy, shown in Fig.~2(a). (For details on the epitaxy and structural and magnetic characterization of the MnTe films, see Methods and Refs.~\cite{Krempasky2024,Amin2024}.)

An earlier study demonstrated a high-resolution imaging of the collinear compensated altermagnetic order in the MnTe epilayers by the X-ray linear and circular dichroism, combined with the photoemission electron microscopy (XMLD and XMCD PEEM) \cite{Amin2024}. Along with the  imaging, the study established a procedure for preparing  single-domain states of the MnTe altermagnet. First, bars of a few-micron width are lithographically patterned, whose crystallographic direction defines the preferred in-plane magnetic easy axis. Next, to select between the two time-reversed states of the given easy-axis, an out-of-plane magnetic field of one or the opposite sign is applied when field-cooling the sample from above the Néel temperature of 310~K. 

Since the post-growth lithography can compromise the sample surface, in the present work we developed  a growth method using a pre-patterned substrate for the molecular-beam epitaxy of MnTe. Using a vacuum suitcase, the patterned MnTe epilayers were then transferred from the growth chamber directly to the synchrotron beam-line chamber without breaking the vacuum. 
In Fig.~2(b), we demonstrate the preparation of a single-domain state of MnTe by XMLD and XMCD PEEM in our pre-patterned sample. The successful angle-resolved photoemission spectroscopy (ARPES) of the pre-patterned sample is illustrated in Fig.~2(c) on the $k_z=0$ momentum-space iso-surface measured at 0.28~eV below the top of the MnTe valence band, using a photon energy of 78~eV. (For more information on the pre-patterned growth, PEEM and ARPES techniques, see Methods.)

Spin-resolved ARPES (SARPES) data on the same iso-surface as in Fig.~2(c) are presented in Fig.~3. Panels in the left column of Fig.~3 show the Néel vector direction vs. the direction of the in-plane momentum scan, and the remaining columns show the corresponding measured spin currents and spin polarization, and the theoretical one-step simulation (for details see Methods) of the momentum-dependent SARPES signal. The three rows of Fig.~3 demonstrate the  key features of the observed phenomenology of the spin polarization on the $k_z=0$ iso-surface: (i) The spin polarization is detected along the $z$-axis, i.e., orthogonal to the Néel vector in accordance with its relativistic spin-orbit coupling origin.  (ii) At opposite momenta, the spin polarization has the same sign, confirming its even-parity and time-reversal-odd nature. (iii) The latter is further confirmed by the observed opposite sign of the spin polarization for the time-reversed states (opposite Néel vectors) of the altermagnet. (iv) The spin polarizations detected in the ${\bf K}_1-\boldsymbol\Gamma - {\bf K}_1$ and ${\bf K}_2-\boldsymbol\Gamma - {\bf K}_2$ scans have an opposite sign, consistent with the alternating sign of the momentum-dependent spin polarization in the ground-state DFT spectra. 

We point out that the experimental comparison between the ${\bf K}_1-\boldsymbol\Gamma - {\bf K}_1$ and ${\bf K}_2-\boldsymbol\Gamma - {\bf K}_2$ scans was realized without rotating the sample,  or without rotating the scanning direction with respect to the fixed orientation of the linear polarization of the vertically-polarized (s-polarization) photon beam. Instead, our pre-patterned sample contained a set of bars with the corresponding magnetic easy axis aligned along the  $[1\bar{1}00]$ crystallographic axis, and another set of bars with the corresponding easy axis rotated by 60$^\circ$. The ${\bf K}_1-\boldsymbol\Gamma - {\bf K}_1$ and ${\bf K}_2-\boldsymbol\Gamma - {\bf K}_2$ scans were measured by focusing the UV photons on the first and on the second set of bars, respectively.

In Fig.~3, the spin-polarization signals are highlighted for momenta with a magnitude larger than $\approx 0.2$~\AA. The spin signal with an antisymmetric component around the $\boldsymbol\Gamma$-point, seen consistently in the experiment and simulations, is a photoemission polarization artifact. This is why the focus in Fig.~3 is on the momenta sufficiently far from the $\boldsymbol\Gamma$-point. Similarly, photoemission artifacts at the top of the valence band led us to perform the spin-polarization measurements 0.28~eV away from the top of the band.  We elaborate in more detail on these key methodology points  in  Fig.~4.

Figs. 4 (a,b) compare experimental ARPES measurements on MnTe epilayers using 78~eV photons with p- and s-polarisation respectively. Whereas for the p-polarised light the bands can be traced close to their extrema, for s-polarisation the spectral weight is suppressed near the top of the valence band. This explains why the UV SARPES measurements in Fig.~3 were performed 0.28~eV below the top of the band.  In addition, we observe an enhanced spectral weight in measurements using s-polarised light around the $\boldsymbol\Gamma$-point. (Note that a similar enhancement near the $\boldsymbol\Gamma$-point is also seen on the iso-surface in Fig.~2(c)). 
The origin of the enhanced intensity can be understood by considering the contribution from different photoemission final-states, as shown by the ARPES simulations using a damped free-electron final state in Fig.~4(c) and a fully scattered, time-reversed low energy electron diffraction (TR-LEED) state \cite{Braun2018} shown in Figs.~4(d-f).  The latter simulations also consistently show the missing contrast near the top of the valence band away from the $\boldsymbol\Gamma$-point. Finally, the SARPES simulations in Fig.~4(e,f) highlight that the final-state effects generate an antisymmetric spin signal concentrated near the $\boldsymbol\Gamma$-point (cf. Figs.~3(d,h,l)). The symmetric signal below the top of the valence band and away from the $\boldsymbol\Gamma$-point, as highlighted in the experimental and theoretical SARPES plots in Fig.~3, is confirmed by the simulations in Fig.~4(e,f) to reflect the spin polarization in the MnTe ground-state spectra.

We now discuss in more details Fig.4 (c-f) in order to illustrate the significance of final-state effects by comparing one-step ARPES simulations for the damped free-electron final state (Fig.~4(c)) with those for the TR-LEED state (Figs.~4(d-f)). The TR-LEED final state consists of a free-electron wave combined with a damped wave field that accounts for phase contributions from scattering sites within the solid \cite{Braun2018}. As a result, TR-LEED states can be used to emphasize the impact of scattering phase shifts on the intensity of the ARPES spectral functions by accounting for the strength of contributions from different atomic orbitals. This results in the modification of the photoelectron's final state wavefunction. The comparison shown in Fig.~4(c-d) demonstrates that the nature of the final state significantly influences the overall intensity. 
This highlights the importance of scattering phase shifts in the accurate ARPES description, and  underscores the necessity of moving beyond simplified free-electron-like models to properly capture ARPES intensity and related phenomena.

To conclude, our observation of the unconventional relativistic spin polarization in the altermagnet enriches the phenomenology of spin-polarized electronic spectra in condensed-matter systems by a fundamentally distinct symmetry type. This may have broad implications in fields including spintronics or topological dissipationless nanoelectonics \cite{Jungwirth2025b,Jungwirth2025a}. Reading and writing information in modern spintronic memories is mediated by spin-polarized electrical currents of a non-relativistic magnetic-ordering origin. In altermagnets, only the $d$-wave magnetic order allows for spin-dependent components of the linear-response conductivity tensor  in the non-relativistic limit. Our MnTe is an example of the numerous  $g$-wave or $i$-wave altermagnets in which the whole non-relativistic conductivity-tensor is spin independent \cite{Smejkal2022a,Jungwirth2025b}. The unconventional relativistic spin polarization, observed in our experiments, reflects a symmetry lowering by the spin-orbit coupling which also enables the electrical current to carry a non-zero spin angular momentum in the higher partial-wave altermagnets.  For electrical currents driven along the $k_z=0$ nodal plane of the MnTe altermagnet, the spin angular momentum carried by the electrical current points along the same $z$-axis as the relativistic spin polarization observed in the electronic spectra, and the sign is reversed by reversing the altermagnetic order \cite{Jungwirth2025b}. In contrast, the conventional odd-parity and time-reversal-even relativistic spin-polarization in normal phases renders the electrical conductivity spin independent.  In the  topological transport, such as the quantum spin Hall effect, the unconventional collinear nature of the relativistic spin polarization in altermagnets is predicted to lead to a  uniquely robust quantization of the spin-Hall conductivity \cite{Jungwirth2025a}. This can facilitate robust non-dissipative charge-to-spin conversion, which so far has been elusive  in conventional topological insulators.

\section{Acknowledgement}
We acknowledge support from Balasubramanian Thiagarajan.
We thank MAX IV Laboratory for time on Beamline BLOCH under proposals 20230293, 20231587, 20241610, 20251165, and Beamline MaxPEEM under proposal 20240976. Research conducted at MAX IV, a Swedish national user facility, is supported by the Swedish Research Council under contract 2018-07152, the Swedish Governmental Agency for Innovation Systems under contract 2018-04969, and Formas under contract 2019-02496. 
D.A.U. acknowledges funding from the SNSF Spark grant CRSK-2228962.
L\v{S} acknowledges funding from the ERC Starting
Grant No. 101165122 and Deutsche Forschungsgemeinschaft (DFG) grant no. TRR 288 - 422213477 (Projects A09 and B05).
S.W.D. and J.M. thank the QM4ST project financed by the Ministry of Education of the Czech Republic grant no. CZ.02.01.01/00/22008/0004572, co-funded by the European Regional Development Fund.
F.G. and J.H.D. acknowledge support from the Swiss National Science Foundation (SNSF) Project No. 200021-200362.
Electron-beam lithography was carried out at the nanoscale and microscale Research Centre, University of Nottingham, supported by EPSRC Grant P/M000583/1. 
O.J.A. acknowledges support from the Leverhulme Trust Grant ECF-2023-755. 
TJ acknowledges support by the Ministry of Education of the Czech Republic CZ.02.01.01/00/22008/0004594 and ERC Advanced Grant no. 101095925.
P.W. acknowledges support from the Royal Society through a University Research Fellowship. The work was supported by the EPSRC grant EP/V031201/1.

%

\newpage

\begin{figure}[h!]
\hspace*{0cm}\epsfig{width=1.0\columnwidth,angle=0,file=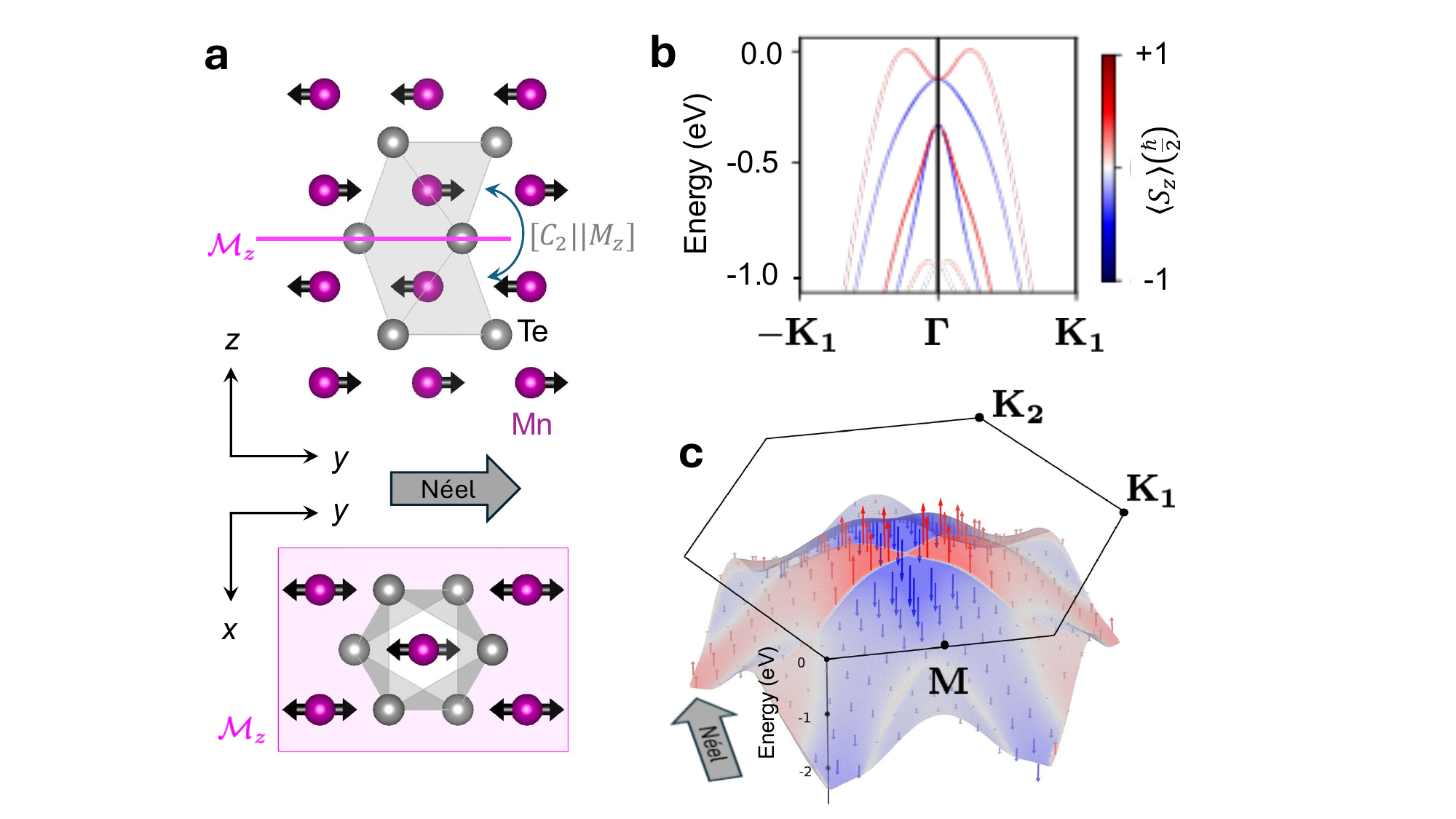}
\caption{Theory of the unconventional relativistic spin polarization in altermagnetic MnTe. (a) Top panel: schematic side-view of the crystal and magnetic structure of MnTe in the $y–z$ plane. $\left[C_{2}||M_{z}\right]$ marks the altermagnetic spin-group symmetry connecting the two opposite Mn sublattices colored in purple, denoted by opposite black arrows, and surrounded by Te octahedra shaded in grey. The relativistic magnetic symmetry group contains mirror plane symmetry $\mathcal{M}_{z}$, marked in magenta. 
Bottom panel: schematic top-view of the crystal and magnetic structure of MnTe in the $x–y$ plane.
(b) Relativistic ab initio band structure of MnTe calculated at the $k_{z}$=0 plane along the $-\bf{K}_{1} - \boldsymbol{\Gamma} - \bf{K}_{1}$ path (for Brillouin zone notation see panel (c)). The Néel vector is oriented along the crystal $y$-axis (see bottom panel (a)). Red - white - blue color bar marks the computed expectation value of the $z$ - component of spin.
(c) Ab initio valence band structure and spin-polarization expectation value calculated at the $k_{z}$=0 plane. In the hexagonal $k_{z}$=0 plane, three wavevectors $\bf{K}_{1}$, $\bf{K}_{2}$, and $\bf{M}$ are highlighted. 
}
\label{f1}
\end{figure}

\begin{figure}[h!]
\hspace*{0cm}\epsfig{width=1\columnwidth,angle=0,file=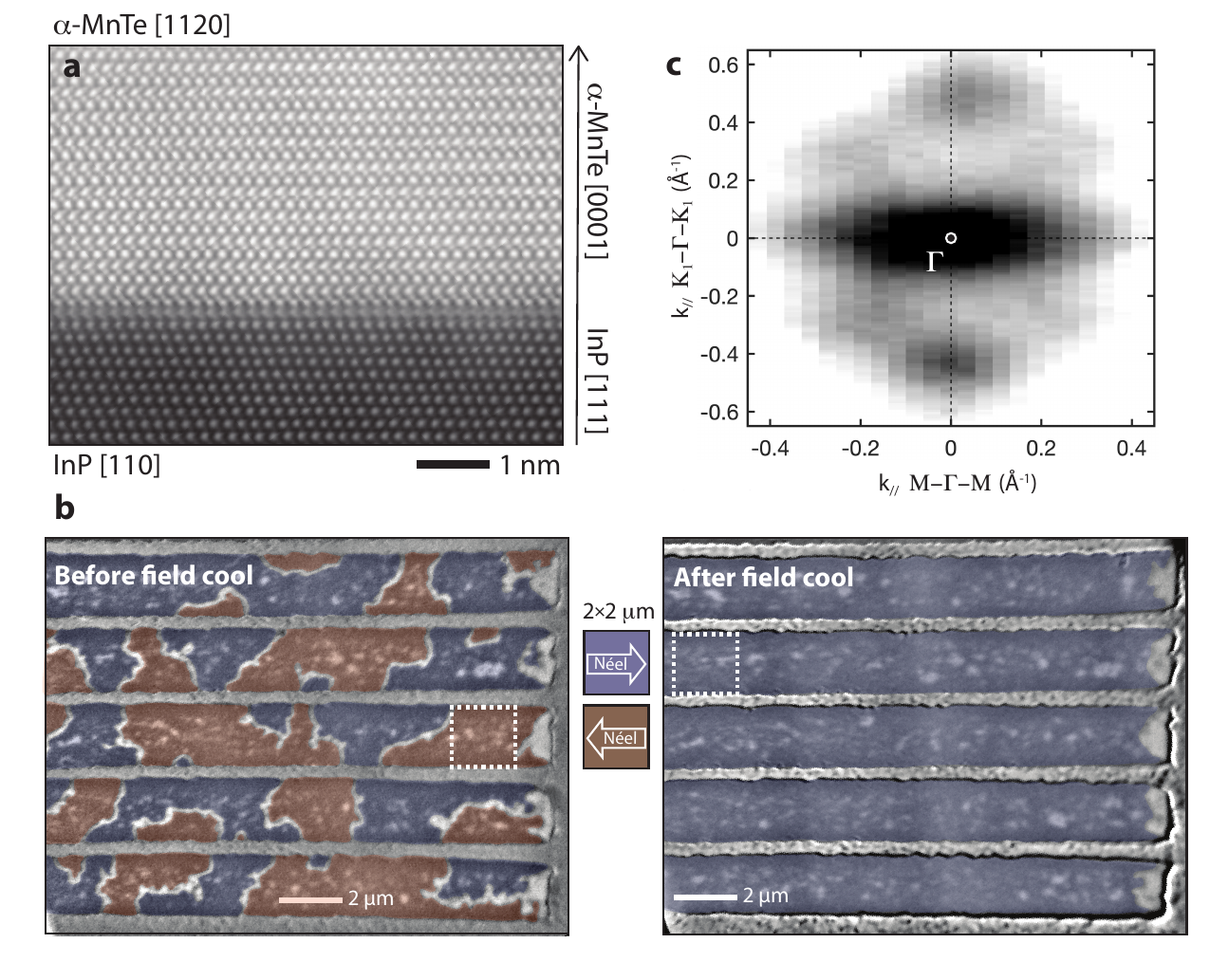}
\caption{Transmission electron microscopy (TEM) imagining and preparation of the single magnetic domain state in the thin films of MnTe. (a) TEM image of the thin films of $\alpha$-MnTe(0001) grown by molecular-beam epitaxy  on an InP(111) substrate. (b) Demonstration of preparation of a single-domain state of MnTe in our pre-patterned sample by X-ray magnetic linear and circular dichroism photoemission electron microscopy. Left panel, before field cooling, and right panel, after field cooling. (c) Angle-resolved photoemission spectroscopy of the pre-patterned sample on the $k_z=0$ momentum-space iso-surface measured at energy 0.28~eV below the top of the MnTe valence band, using a photon energy of 78~eV. 
}
\label{f2}
\end{figure}

\begin{figure}[h!]
\hspace*{0cm}\epsfig{width=.9\columnwidth,angle=0,file=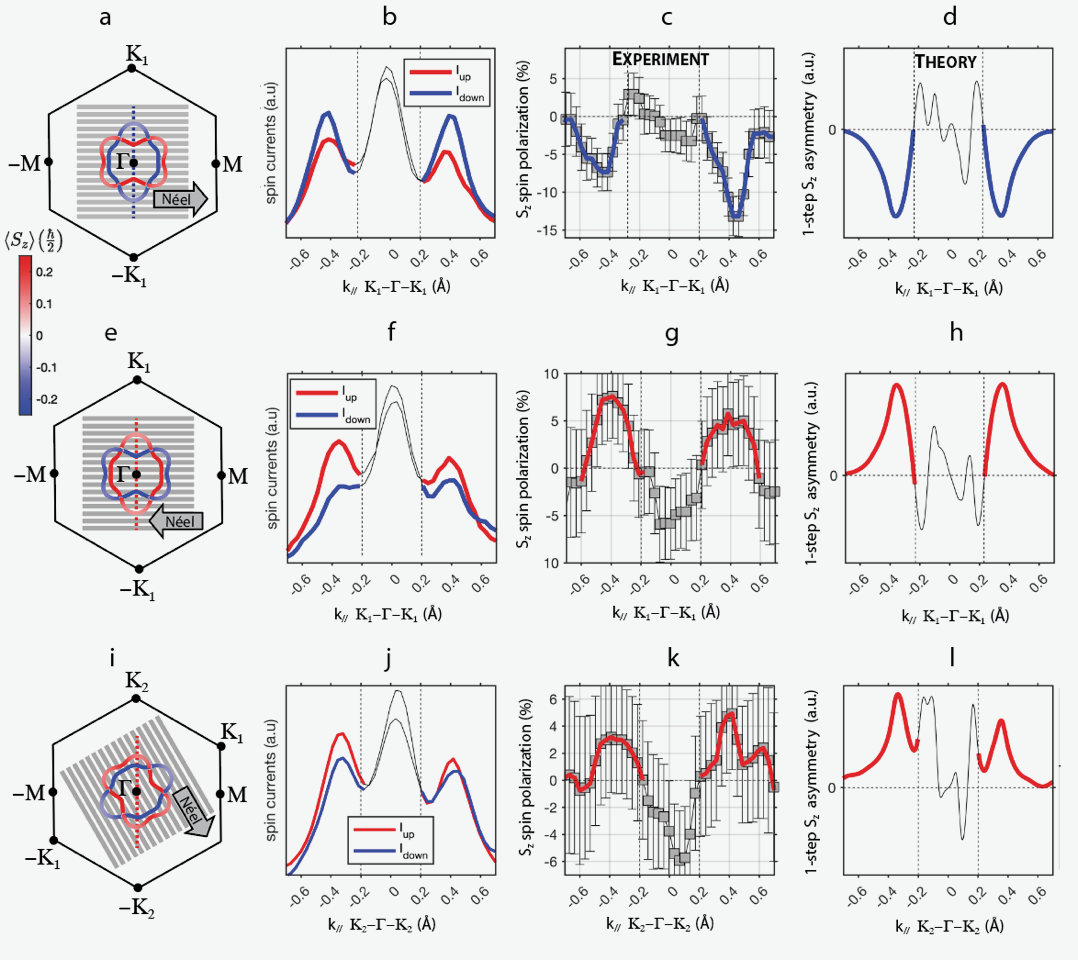}
\caption{Spin-resolved and angle-resolved photoemission spectroscopy (SARPES) observation and one-step photoemission theory of unconventional out-of-plane spin polarization $S_{z}$. The $S_{z}$ sign reversal under 180$^\circ$, and 60$^\circ$ rotation of the Néel vector, respectively, demonstrates the unconventional out-of-plane, time-reversal broken and alternating nature of the spin polarization. 
The first column (a, e, i) shows three experimental geometries with the orientation of the patterned sample (patterning direction marked by grey lines), the Néel vector, and the ab initio calculated constant-energy isosurface at 0.28 eV below the top of the valence band.
The second column (b, f, j) presents the SARPES spin-current experimental data, and the third column (c, g, k) shows the corresponding experimental out-of-plane $S_z$ spin polarization.
The last column (d, h, l) displays one-step photoemission calculations of the out-of-plane spin-polarization asymmetry, which are consistent with the experimental results.
The first two rows demonstrate the reversal of the spin-polarization sign along the  $-\bf{K}_{1} - \boldsymbol{\Gamma} - \bf{K}_{1}$ path upon reversing the Néel vector direction, while the last row shows the reversal when measuring along the  $-\bf{K}_{2} - \boldsymbol{\Gamma} - \bf{K}_{2}$ direction.}
\label{f3}
\end{figure}

\begin{figure}[h!]
\hspace*{0cm}\epsfig{width=.75\columnwidth,angle=0,file=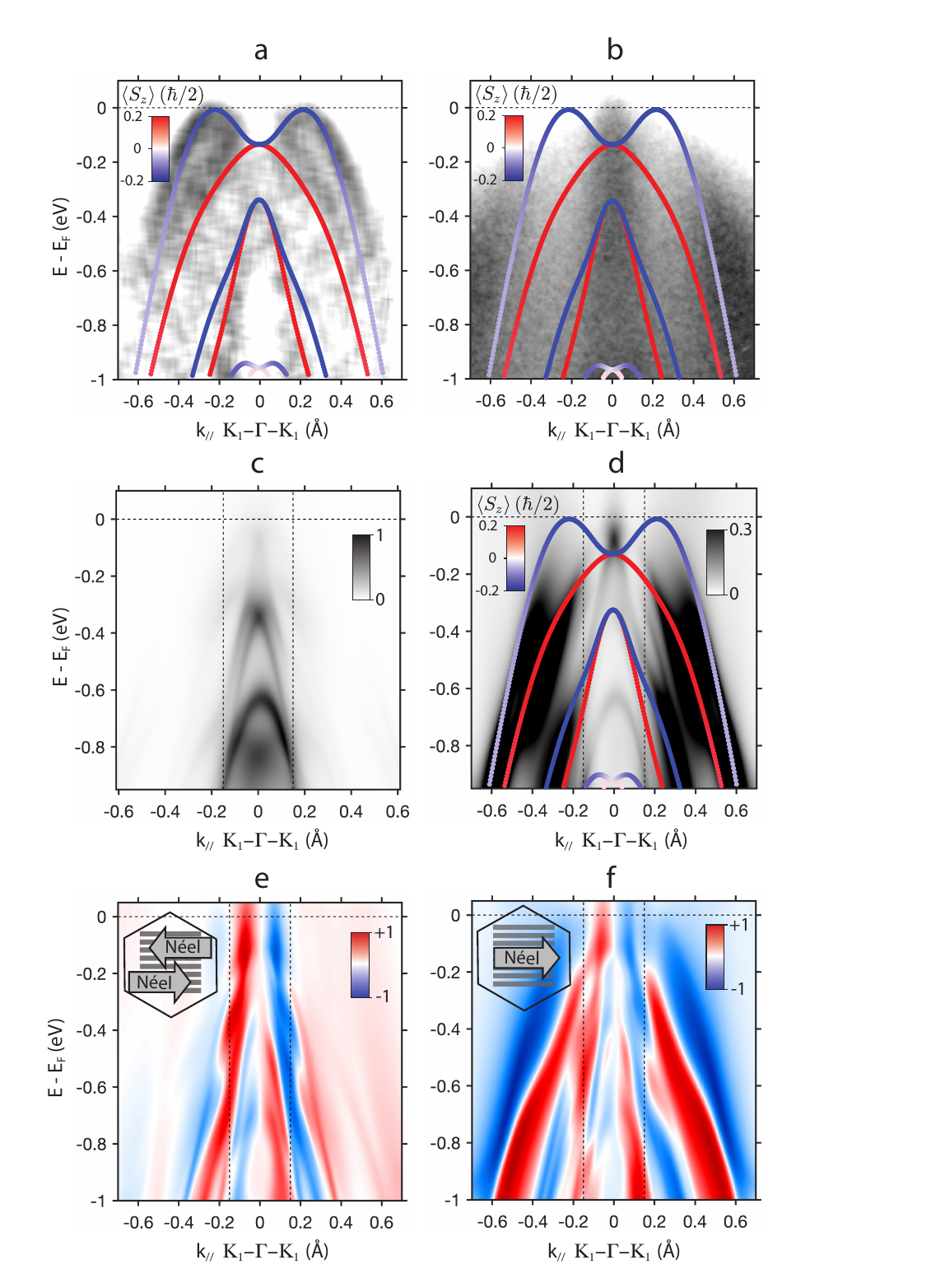}
\caption{The ARPES measurement and one-step photoemission theory of band maps along  $-\bf{K}_{1} - \boldsymbol{\Gamma} - \bf{K}_{1}$ wavevector path. Measured ARPES band maps with a p-polarized (a) and s- (b) polarized photons, overlayed with ab initio spin-polarized band structure. 
(c,d) One-step photoemission calculation of the intensity considering photoemission final states in the form of (c) free-electron like and (d) time-reversed low energy electron diffraction (TR-LEED).
(e,f) One-step photoemission calculations with TR-LEED final states for (e) the average of the two time-reversed domains and (f) one domain only, as indicated in the insets. 
}
\label{f4}
\end{figure}

\end{document}